\documentclass[longauth]{aa} 

\usepackage{graphicx}
\usepackage[colorlinks=true,citecolor=blue]{hyperref}
\usepackage{soul}
\usepackage{xcolor}

\newcommand{\MJup}{\ensuremath{M_{\mathrm{Jup}}}\xspace}

\newcommand{\MSun}{\ensuremath{M_{\odot}}\xspace}

\newcommand{\LSun}{\ensuremath{L_{\odot}}\xspace}

\newcommand{\lsd}
\begin{document}

\title{The environment around young eruptive stars}
   
\subtitle{SPHERE/IRDIS polarimetric imaging of 7 protostars\thanks{This publication made use of data from the ESO programs 098.C-0422(A), 099.C-0147(B), 0100.C-0408(A), and 1104.C-0415(D)}}

   \author{A. Zurlo\inst{1,2}, P. Weber\inst{3,2,4}, S. P\'erez\inst{3,2,4}, L. Cieza\inst{1,2}, C. Ginski\inst{5}, R.G. van Holstein\inst{6}, D. Principe\inst{7}, A. Garufi\inst{8}, A. Hales\inst{9}, J. Kastner\inst{10}, E. Rigliaco\inst{11}, G. Ruane\inst{12}, M. Benisty\inst{13}, C. Manara\inst{14}}
         
   \institute{\inst{1} Instituto de Estudios Astrof\'isicos, Facultad de Ingenier\'ia y Ciencias, Universidad Diego Portales, Av. Ej\'ercito 441, Santiago, Chile \\
\inst{2} Millennium Nucleus on Young Exoplanets and their Moons (YEMS), Santiago, Chile \\
              \email{alice.zurlo@mail.udp.cl} \\
              \inst{3} Departamento de Física, Universidad de Santiago de Chile, Av. Victor Jara 3659, Santiago, Chile \\
              \inst{4} Center for Interdisciplinary Research in Astrophysics and Space Science (CIRAS), Universidad de Santiago de Chile\\
              \inst{5} School of Natural Sciences, University of Galway, University Road, H91 TK33 Galway, Ireland \\
              \inst{6} European Southern Observatory, Alonso de Córdova 3107, Casilla 19001, Vitacura, Santiago, Chile \\
             \inst{7} Massachusetts Institute of Technology, Cambridge, MA, USA \\
            \inst{8} INAF, Osservatorio Astrofisico di Arcetri, Largo Enrico Fermi 5, 50125 Firenze, Italy \\
 \inst{9} National Radio Astronomy Observatory, 520 Edgemont Road, Charlottesville, VA 22903-2475, USA \\
 \inst{10} Center for Imaging Science and Laboratory for Multiwavelength Astrophysics, Rochester Institute of Technology, Rochester, NY 14623, USA \\
 \inst{11} INAF-Osservatorio Astronomico di Padova, Vicolo dell'Osservatorio 5, Padova, Italy, 35122-I \\
 \inst{12} Jet Propulsion Laboratory, California Institute of Technology, 4800 Oak Grove Dr., Pasadena, CA91109, USA \\
            \inst{13} Université Côte d’Azur, Observatoire de la Côte d’Azur, CNRS, Laboratoire Lagrange, Bd de l’Observatoire, CS 34229, 06304 Nice cedex 4, France \\
            \inst{14} European Southern Observatory, Karl-Schwarzschild-Strasse 2, 85748, Garching bei München, Germany
}

   \date{Accepted 14 March 2024 }
  \titlerunning{The environment around young eruptive stars}
\authorrunning{A. Zurlo et al.}

  \abstract
   {}
   {Eruptive stars are a class of young stellar objects that show an {abrupt} increase in their luminosity. These burst-like episodes are thought to dominate the stellar accretion process during the class 0/class I stage. We present {an overview of }a survey of seven episodically accreting protostars aimed at studying their potentially complex circumstellar surroundings.}
   {The observations were performed with the instrument SPHERE, mounted at the Very Large Telescope. SPHERE is equipped with an extreme adaptive optics system that allows high-contrast imaging. We observed the eruptive stars in $H$-band with the near-infrared imager IRDIS and used the polarimeter to extract the polarized light scattered from the stars' surroundings.}
  {{We produced polarized light images for three FUor objects, Z~CMa, V960~Mon, and FU~Ori, and four EXor objects, XZ~Tau, UZ~Tau, NY~Ori, and EX~Lup. We calculated the intrinsic polarization fraction for all the observed stars.}
  {In all systems we registered scattered light from around the primary star.} 
  FU~Ori and V960~Mon are surrounded by complex structures including spiral-like features.
  In Z~CMa, we detected a point source 0\farcs7 to the northeast of the primary. Based on the astrometric measurements from archival Keck/NIRC2 data, we find this source to be a third member of the system.
  Further, Z~CMa displays an outflow extending thousands of au.
  Unlike the other EXor objects in our sample, XZ~Tau shows bright, extended scattered light structures, also associated with an outflow on a scale of hundreds of au. 
  The other EXors show relatively faint disk-like structures in the immediate vicinity of the coronagraph.}
   {Each object shows a unique environment, but we classified the 7 objects into 3 categories: systems with illuminated outflows, asymmetric arms, and faint disks. Asymmetric arms were only found around FUor objects, while faint disks seem to predominantly occur around EXors. Importantly, for Z~CMa the detection of the faint extended structure questions previous interpretations of the system's dynamic state. {The streamer which was associated with a fly-by object turned out to be part of a huge outflow extending 6000 au. }}


   \maketitle
%

\section{Introduction}
Young eruptive stars are a class of young stellar objects (YSO) that display sudden and intense increases in their luminosity, followed by a gradual decline over a period of weeks to years. The light curves and spectra of these objects exhibit a wide range of variations, making their classification a challenging task \citep[see, e.g.,][]{2008AJ....135.1421G}. {Historical studies on the optical nebulosity around these objects underscore the ubiquity of reflection nebulae found in association with these eruptive stars at sub parsec scales \citep[see review by][]{1996ARA&A..34..207H}.} 

Within the group of eruptive stars, two loosely defined categories are typically recognized depending on the duration of the outburst and the presence of certain spectral features: FU~Ori objects, often referred to as FUors, named after the class prototype, FU~Orionis; and EXor objects, named after the prototype EX~Lup \citep[see][for a recent discussion on classification schemes]{2023ASPC..534..355F}. {FUors have outbursts with sustained increases in bolometric luminosity, and their decay timescale is about 100 years, their spectrum is dominated by the disk viscous heating. On the other hand, EXors show spectra similar to T Tauri stars with high accretion. Their outbursts have amplitudes of several magnitudes, last for months, and can recur after a few years \citep{2023ASPC..534..355F}.} 
These eruptive stars are believed to go through enhanced episodic accretion \citep[see][for a review]{2014prpl.conf..387A, 2019ApJ...884..146T,2021ApJS..256...30K}, a process that could {explain the wide range of protostellar luminosities and} solve the so-called "luminosity problem", which refers to the discrepancy between the observed luminosities of young, embedded low-mass protostars and the predicted luminosities from standard theoretical models \citep[e.g.,][]{1990AJ.....99..869K}. Both FUors and EXors objects undergo periodic bursts, in which material falls onto the disk from the circumstellar envelope. Therefore, resolved maps of the surroundings of these eruptive stars can help us reconstruct the accretion history of the source. These outbursts can last for months (in the case of EXors) or even years/decades (FUors) and are believed to be caused by a variety of mechanisms which can be investigated through high-resolution imaging, including gravitational instabilities in the disk \citep{2001MNRAS.324..705A, 2009ApJ...701..620Z}, thermal instability \citep{1994ApJ...427..987B}, disk fragmentation \citep{2005ApJ...633L.137V, 2012ApJ...746..110Z}, or interactions with a companion star or sub-stellar object \citep{1992ApJ...401L..31B,2004MNRAS.353..841L}. 

The environment around eruptive stars has been studied intensively in the (sub-)millimeter. Recent Atacama Large Millimeter/submillimeter Array (ALMA) observations revealed complex structures in the environment around several FUor/EXor objects. In the disk of V883~Ori, for the first time, the water snow line has been detected \citep{2016Natur.535..258C}, and the same object presents a wide outflow in the gas emission \citep{2017MNRAS.468.3266R,2022MNRAS.515.2646R,2023Natur.615..227T}. The FUor V2775~Ori shows a spectacular hourglass shape in gas emission \citep{2017MNRAS.465..834Z}. The object HBC~494 is surrounded by a very wide outflow \citep{2017MNRAS.466.3519R} and has a resolved close companion that also hosts a compact disk \citep{2023MNRAS.523.4970N}. V1647~Ori shows misaligned outflow with respect to its host nebula \citep{2018MNRAS.473..879P}. Last but not least, the Z CMa system's outflow was first identified by \cite{1960ApJS....4..337H} with several nebulosities found associated with this active system since then \citep{2023Galax..11...64L}.

The ALMA continuum emission of several of these sources is presented in recent works that include radiative transfer modeling to compute the disk masses. For example,  \cite{2018MNRAS.474.4347C} present the analysis of the dust continuum emission and the masses of three FUor, four EXor, and the borderline object V1647 Ori. The inferred disk radii are smaller than T~Tauri stars with comparable disk masses, suggesting that disks around outbursting objects are more compact than around regular stars. \citet{2020ApJ...900....7H} presented an ALMA survey of 4 objects at 1.3 mm and 0\farcs4 resolution: two FUors (V582 Aur and V900 Mon), one EXor (UZ~Tau~E), and one source with unambiguous FU/EXor classification (GM~Cha). From the continuum emission, they derived the masses of the dust in the disks and found that FUors are more massive than Class~0, I, and II of similar sizes. EXors have masses and radii comparable to disks around Class~II objects. In a more recent survey, \citet{2021ApJS..256...30K} conducted an ALMA 1.3 mm continuum survey of 10 FUors and FUor-like objects. Their findings suggest that the disks of FUors are typically a factor of 3-4 more massive and a factor of 2-5 smaller in size than class II and class I objects, and that a good fraction of them might be gravitationally unstable. 

A compact radius in the disk size might also be related to the presence of a stellar companion, as demonstrated in several studies \citep[e.g.,][for a review]{2019ApJ...882...49L, 2019ApJ...872..158A, 2023EPJP..138..411Z}. In general, disks observed in the (sub)-mm around multiple stars are more compact than the ones around single stars. Even if a fraction of eruptive stars are in multiple systems, this might not necessarily mean that their disks have smaller sizes because of the multiplicity.

Fewer studies on the environment of eruptive stars have been carried out with AO-assisted observations in the near-IR. At these wavelengths, the most successful tool to directly image the immediate surroundings of bright stars is the employment of a polarimeter to discern stellar light and light scattered off by circumstellar material \citep{2023ASPC..534..605B}. With 8-m class telescopes such as the Very Large Telescope (VLT), one can obtain resolved maps of the polarized light at separations larger than 0\farcs1. Any polarimetric signal enclosed in the region closer in will be contained in the unresolved polarization of the central beam, which can be now measured for observations done with the Spectro-Polarimetric High-contrast Exoplanet REsearch (SPHERE) at the VLT in Chile \citep{vanHolstein2020}.

{Polarimetric imaging is a potent tool for gaining a better understanding of the environment around young stars, revealing details about their composition, structure, and dynamics \citep[see, for example,][]{2016SciA....2E0875L,2018ApJ...863...44A}.  It also aids in identifying disk misalignments via shadows \citep[see, for example,][]{2015ApJ...798L..44M, 2023MNRAS.526.2077A}, and detecting magnetically aligned grains \citep{2022AJ....164...99Y}. For a detailed overview of scattered and polarized light observations of protoplanetary disks, we direct readers to \citet{2023ASPC..534..605B}, and references therein.} 

In the context of multiple stellar systems, polarimetric observations can additionally uncover details about the illumination pattern \citep{1993ApJ...411..767W,2023MNRAS.518.5620W}. This capability enables the identification of asymmetries and complex features, shedding light on the system's geometry and helping us discern which star is illuminating specific dust regions \citep{2023MNRAS.518.5620W}. Such an approach enriches our comprehension of the early phases of star formation and contributes to our knowledge of the processes involved in planet formation, as well as the various conditions found in young stellar neighborhoods.

At the time of writing, there are about a hundred or so known  {young} objects that {are classified} as eruptive stars.
This number includes the $\sim$40 objects presented in \cite{2014prpl.conf..387A} {in a volume of 1700 pc} and 80 outbursting VVV candidates with eruptive classification confirmed spectroscopically {in a volume of 10 kpc} \citep{2017MNRAS.465.3039C, 2021MNRAS.504..830G}.
Most of these objects are young ($<$1 Myr) pre-main sequence stars and still heavily embedded in material from their nascent molecular cloud or parent envelope, or very distant in the case of most of the VVV sample, which {hinders} their observability with wavefront sensors in the optical.
Pioneering work with the High Contrast Instrument for the Subaru Next Generation Adaptive Optics (HiCIAO) allowed to reveal tantalizing details of complex structures around four of these objects in the near-IR \citep[see,][]{2016SciA....2E0875L}. 

Here we present a near-infrared ($H$-band) survey of 7 EXors and FUors performed with VLT/SPHERE. The SPHERE adaptive optics system operates in the optical and thus most FUor/EXors suffer too much extinction to be observable with this instrument. The sample in this paper consists of seven well-studied examples of eruptive YSOs that are particularly amenable to observations by SPHERE. The manuscript is organized as follows: in Sec.~\ref{s:obs} we present the observations and data reduction of the sources, in Sec.~\ref{s:res} we present the results and a short discussion on each individual target, and in Sec.~\ref{s:disc} we put the main results of our survey into context. Finally, Sec.~\ref{s:con} summarizes our conclusions.

\section{Observations and data reduction}
\label{s:obs}
The data presented in this manuscript are part of the survey under ESO programs 098.C-0422(A) and 0100.C-0408(A) (PI: Cieza). {The original SPHERE survey included all the FUors and EXors presented in \cite{2014prpl.conf..387A} with a R mag brighter than 13, which is the SPHERE limit for the adaptive optics: Z~CMa, XZ~Tau, FU~Ori, V960~Mon, UZ~Tau, NY~Ori. {EX~Lup} was a protected target of the SPHERE GTO, and their data are now available in the archive under ESO program 099.C-0147(B) (PI: Beuzit). As the quality of the observations for Z CMa was such that no polarimetric cycles were completed, we added archival data from 1104.C-0415(D) (PI: Ginski\footnote{These data are part of the ESO large program called Disk Evolution Study Through Imaging of Nearby Young Stars \citep[DESTINYS;][]{2020A&A...642A.119G} led by C. Ginski.}).  }.

 The objects XZ~Tau, UZ~Tau, and NY~Ori are categorized as EXors in \cite{2022ApJ...929..129G}, while Z~CMa and V960~Mon are bona fide FUors \citep{2023ApJ...945...80C}. The FU~Ori and EX~Lup systems are the prototypes of the two categories. The survey was performed with the SPHERE instrument \citep{2019A&A...631A.155B}, a high-contrast imager installed at the VLT and equipped with an extreme adaptive optics system \citep[SAXO;][]{Fusco:06, 2014SPIE.9148E..0OP}. The near-infrared (NIR) arm is composed of the infrared dual-band imager and spectrograph \citep[IRDIS;][]{Do08} and an integral field spectrograph \citep[IFS;][]{Cl08}. For this study, we used the IRDIS linear-polarimetric imaging mode, to perform the Polarimetric Differential Imaging \citep[PDI;][]{deBoer2020,vanHolstein2020}. {The observations were performed in H-band, with an angular resolution of 0\farcs05 and Strehl ratio of 90\% for good seeing conditions {($<$ 0\farcs8)}\footnote{From the SPHERE manual}.} {For all the observations the N\_ALC\_YJH\_S coronagraph was implied }\citep[185 mas diameter,][]{2011ExA....30...59G}.

The observations are summarized in Table~\ref{t:obs}. 
The observations were performed in field-stabilized mode (all but Z CMa's), so the contrast is not optimal for the search for companions. {When observing single protoplanetary disks that exhibit a high degree of axisymmetry, it is possible to separate the signal into Qphi and Uphi components. This decomposition, coupled with the assumption of single scattering events, enables the isolation of the majority of noise within the Uphi component, with the exception of cross-talk contributions \citep[see, e.g.,][]{2018ApJ...863...44A}. However, in the context of our observations, which encompass complex, extended signals potentially illuminated by multiple stars within the field, such decomposition cannot be performed. As an alternative, we present the rms noise of the background in the PI images as an approximation of the noise levels. It is important to note that these values are relatively high, considering that the PI images are generated by summing the squares of Q and U (see below). The rms of the background in our observations spans from 0.006 mJy/arcsec$^2$ in the best case to 0.1  mJy/arcsec$^2$ in the worst (see Fig.~\ref{f:appendix}).}

The objects were observed on four different nights. Both nights of 2017-10-12 and 2017-10-27 had in general exquisite atmospheric conditions during the whole observing time. The seeing value spanned from a minimum of 0\farcs3 to a maximum of 0\farcs8. The wind speed was lower than 7$\,$ms$^{-1}$ during both nights, {the coherence time} was lower than 5$\,$ms. During those two nights, the targets observed were XZ~Tau, UZ~Tau, and FU~Ori. The observations were performed in delegated visitor mode.  

\begin{table}
\begin{minipage}{0.5\textwidth}
\caption{Summary of the $H$-band observations of the SPHERE/IRDIS survey of eruptive stars. DIT gives the detector integration time per frame, \#frames is the number of frames, $\tau_{\rm int}$ the total on-target integration time.} 
\label{t:obs}
\renewcommand{\footnoterule}{}  
\centering
\begin{tabular}{lcccc}
\hline
\hline
  Target    & Night         & DIT[s] & \#frames & $\tau_{\rm int}$[min]\\
           \hline\\[-0.35cm]
            Z CMa${\footnote{from the DESTINYS survey (PI: Ginski)}}$ &  2022-12-17& 32   & 104 & 55.5 \\
            XZ Tau & 2017-10-12  & 64 & 16 & 17.1 \\
            FU Ori & 2017-10-12  & 24 & 24 & 9.6\\
            V960 Mon & 2016-12-17 & 64 & 16 & 17.1\\
            UZ Tau & 2017-10-27  & 64 &14 & 14.9 \\
            NY Ori &  2016-12-16 & 32& 24 & 12.8  \\[-0.1cm]
            EX Lup${\footnote{from the SPHERE GTO survey (PI: Beuzit)}}$ & 2017-05-16  & 64& 22 &  23.5 \\
\hline
\end{tabular}
\end{minipage}
\end{table}

To process the data we used the IRDIS Data reduction for Accurate Polarimetry reduction pipeline \citep[IRDAP, version 1.3.4;][]{vanHolstein2020} and retrieved the images of the Stokes $Q$ and $U$ components of the linear polarization. 
IRDAP uses a data-independent model of the instrumental polarization and polarization cross-talk to correct the observed polarimetric frames. This allows us to measure the spatially unresolved polarization of the stars. The systems presented in this study contain multiple stars within the field-of-view (FoV) {of 12\farcs0$\times$ 12\farcs0}. The unresolved stellar polarization is carried in the corresponding stellar halo and can contaminate the scattering regions of interest. 

Finally, we combine $Q$ and $U$ components to recover the intensity of linear polarization:
\begin{equation}
    PI = \sqrt{Q^2+U^2}\,.
\end{equation}
\label{eq:1}
In addition, IRDAP computes images of $Q_\phi$ and $U_\phi$ \citep{2006A&A...452..657S} using the definitions of \cite{deBoer2020}. The $Q_\phi$ image is typically preferred for displaying circumstellar environments in polarized light as it has lower noise in comparison to $\mathit{PI}$. However, in the presence of multiple light sources, the interpretation of the $Q_\phi$ and $U_\phi$ images is non-trivial \citep{2023ApJ...952L..17W}. We use the $PI$ image for Z CMa, XZ Tau, FU Ori, and V960 Mon because for these objects there are multiple light sources and the circumstellar disks are bright. On the other hand, for UZ Tau, NY Ori, and EX Lup we use the $Q_\phi$ image because the circumstellar material around these objects is faint and there is only a single, central light source.

To reduce the presence of Poisson-noise in our images, we further made use of the {\tt denoise}-package\footnote{\url{https://github.com/danieljprice/denoise}} which is based on the particle smoothing kernel described within the context of the visualization tool {\tt SPLASH} \citep{2007PASA...24..159P}.
The denoising procedure is most efficient when directly applied to the $Q$ and $U$ products before combining them to $PI$ or $Q_\phi$.
It has to be applied with care, however, in regions where $Q$ and $U$ change strongly over small spatial scales, as the smoothing can lead to artificial cancellation of polarization.
This is typically the case in the bright regions close to the image center.
We, therefore, apply a moderate smoothing kernel ({\tt -{}-beam=1.0}) and refrain from modifying the bright areas (where the signal-to-noise is high anyway) by setting an individually selected maximum value ({\tt -{}-imax}) for the procedure.

\section{Results}
\label{s:res}

In the following sections, each object that is part of this survey is presented individually. We summarize the data present in the literature for each individual target and the new results derived from the SPHERE observations. A general discussion follows the dedicated Sections for each target. 


\begin{figure*}[h!]
\begin{center}
\includegraphics[width=0.9\textwidth]{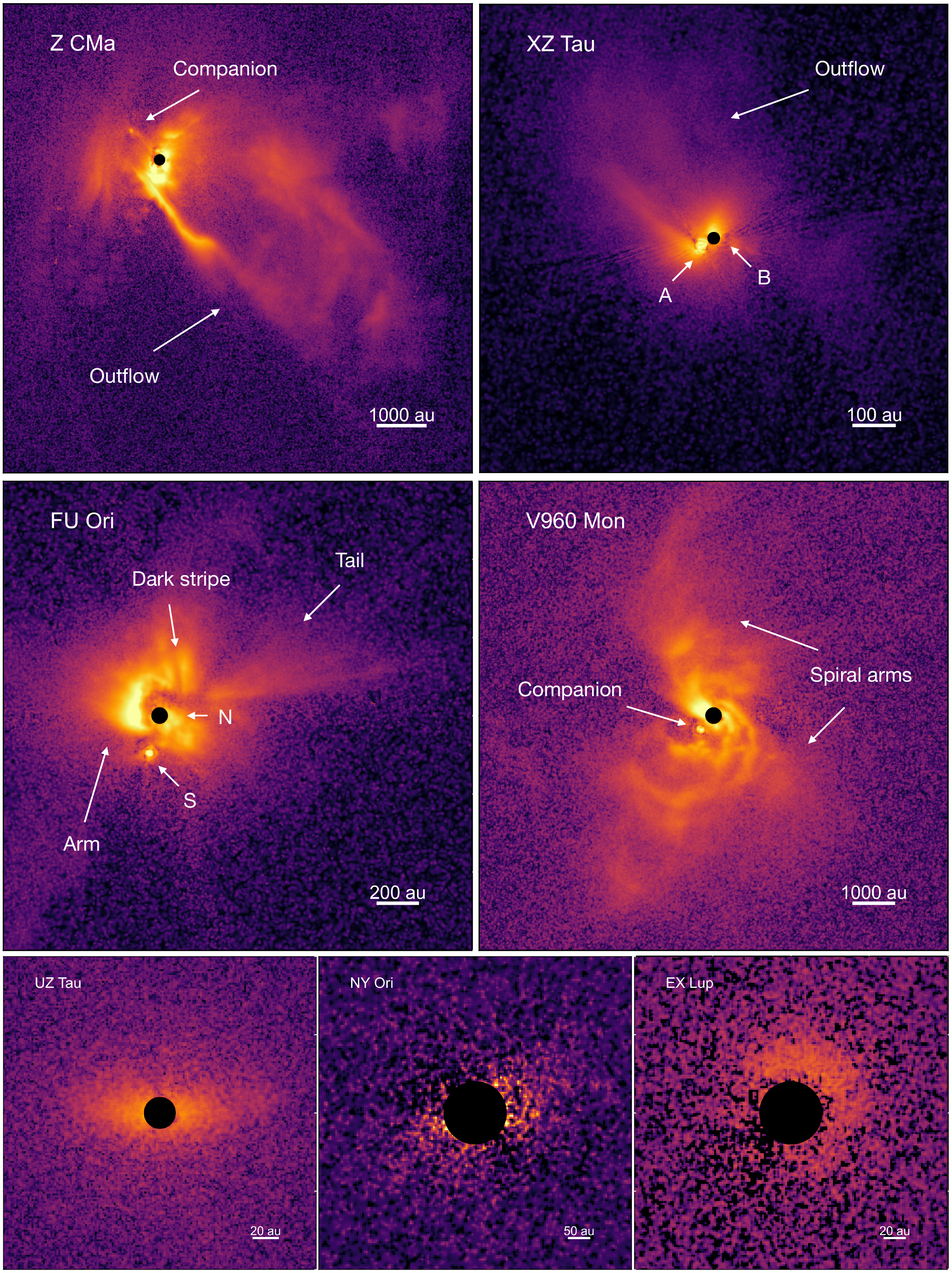}
  \caption{Gallery of all the SPHERE polarized light images of the eruptive stars in $H$ band. North is up, East is left. The black circle indicates the region masked by the coronagraph. {The color scale was optimized for a better visualization, in Fig.~\ref{f:appendix} a color bar indicates the values of the surface brightness.} }
\label{f:mos}
\end{center}
\end{figure*}

\subsection{Z CMa}
Z Canis Majoris is a young binary star composed of an FU Ori-like object, the southeast component \citep[e.g.,][]{2023A&A...672A.158S}, and an EXor object, the northern star \citep{2004MNRAS.349.1516V, 2017A&A...597A..91B}{, which is also a Herbig Be star}. The two stars, separated by 0\farcs1, were resolved via high-contrast imaging several times in the past \citep[see, e.g.,][]{2013ApJ...763L...9H,2015A&A...578L...1C,2016A&A...593L..13A, 2017A&A...597A..91B,2018ApJ...864...20T}. \citet{2016A&A...593L..13A} presented detection {of} both companions, which prevents the association of the extended features around the binary star to one particular star. We assume as a distance the value proposed in \citet{2022NatAs...6..331D} of $1125 \pm 30$ pc, obtained by calculating the {Gaia EDR3} distance to 62 young members of CMa R1. 

\begin{figure}[h!]
\begin{center}
  \includegraphics[height=0.5\textwidth]{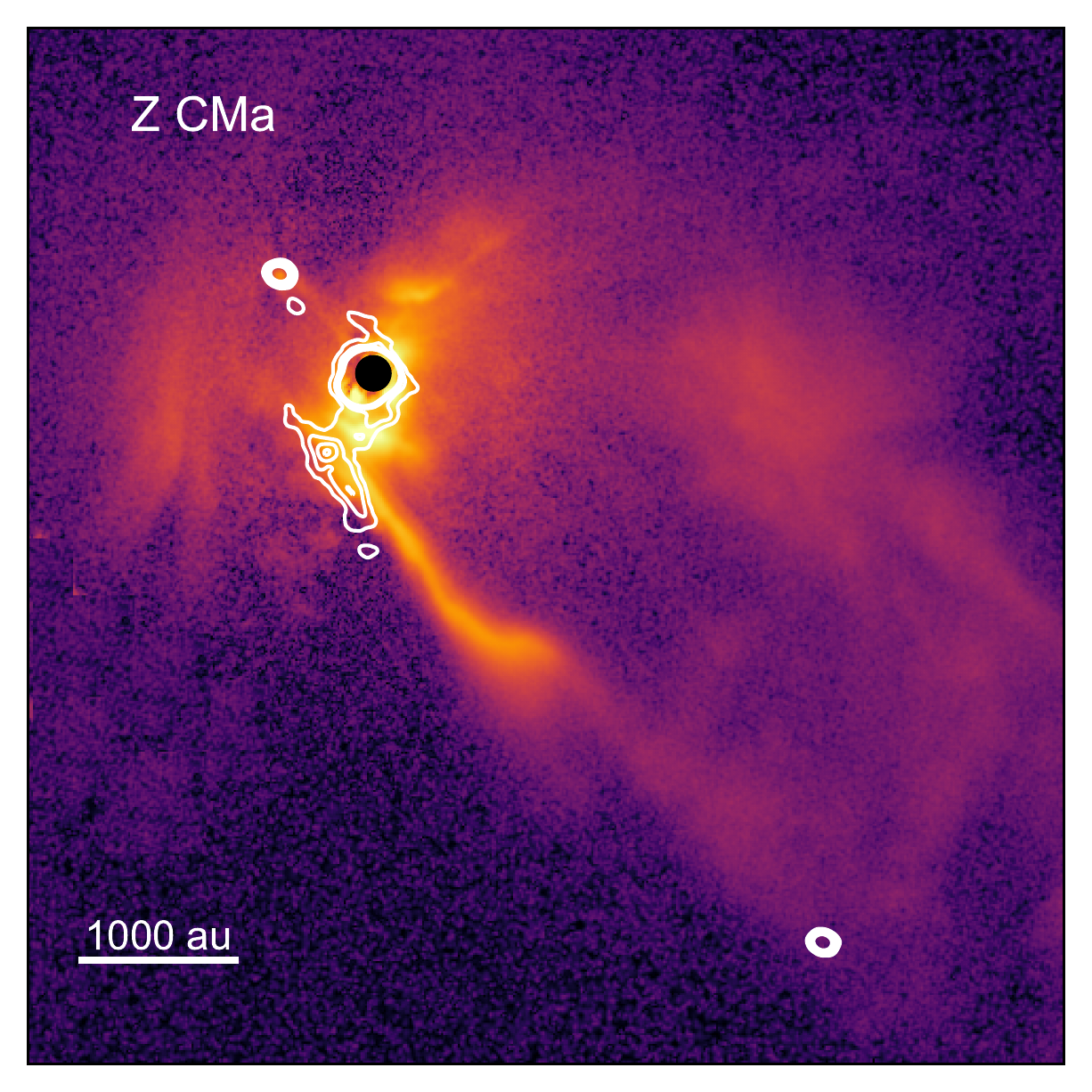}
  \caption{IRDIS image of Z CMa with the ALMA 1.3 mm continuum from \citet{2022NatAs...6..331D} contours overplotted in white. The stellar companion is detected in the ALMA image, and the object at the South-West of the star \citep[labeled emission ``C'' in][]{2022NatAs...6..331D} has no NIR counterpart. }
\label{f:super}
\end{center}
\end{figure}

Scattered light images revealed a large disk around the system, extending 1500 au
\citep{2016SciA....2E0875L}. In the millimeter, on the other hand, ALMA detected compact emission ($\sim$ 20 au) around each of the two components \citep{2022NatAs...6..331D}. But the most interesting feature, revealed in both the NIR and millimeter is the enormous streamer towards the south-west {, as shown in Fig.~\ref{f:super}}. One scenario proposed to explain the origin of this streamer is the encounter with a fly-by object \citep{2022NatAs...6..331D}. Hydrodynamical simulations of a fly-by object interacting with the material around the two stars and subsequent radiative transfer showed that such an event can reproduce the observations of the streamer. Moreover, an unresolved compact source emission in the ALMA images is consistent with being the perturber fly-by. This {putative} perturber is not detected in the near-infrared. The fact that the object is not detected in the SPHERE images means that its mass is below our detection limits. If the radio signal is real {(SNR $\sim$ 27)}, the object would be most probably substellar or a distant background object. Deeper and more sensitive observations are needed to confirm its nature. 

The final Z~CMa image is shown in Fig.~\ref{f:mos}. The polarized light image shows several faint structures that were never detected before. The most spectacular structure is an arch-shaped filament that seemingly connects the streamer, extends for thousands of au, and {ends on the central star}. This detection suggests that the streamer is in reality a portion of an outflow that is brighter {due to} forward scattering. Opposing this arch, we also find a dim contribution east of the star, {which} appears to be substructured, potentially shaped by shock-waves connected to the outburst event. {The SPHERE image further reveals light reflected from the near side of the outflow, northwest of the central star.} 

\begin{figure}[h!]
\begin{center}
  \includegraphics[height=0.45\textwidth]{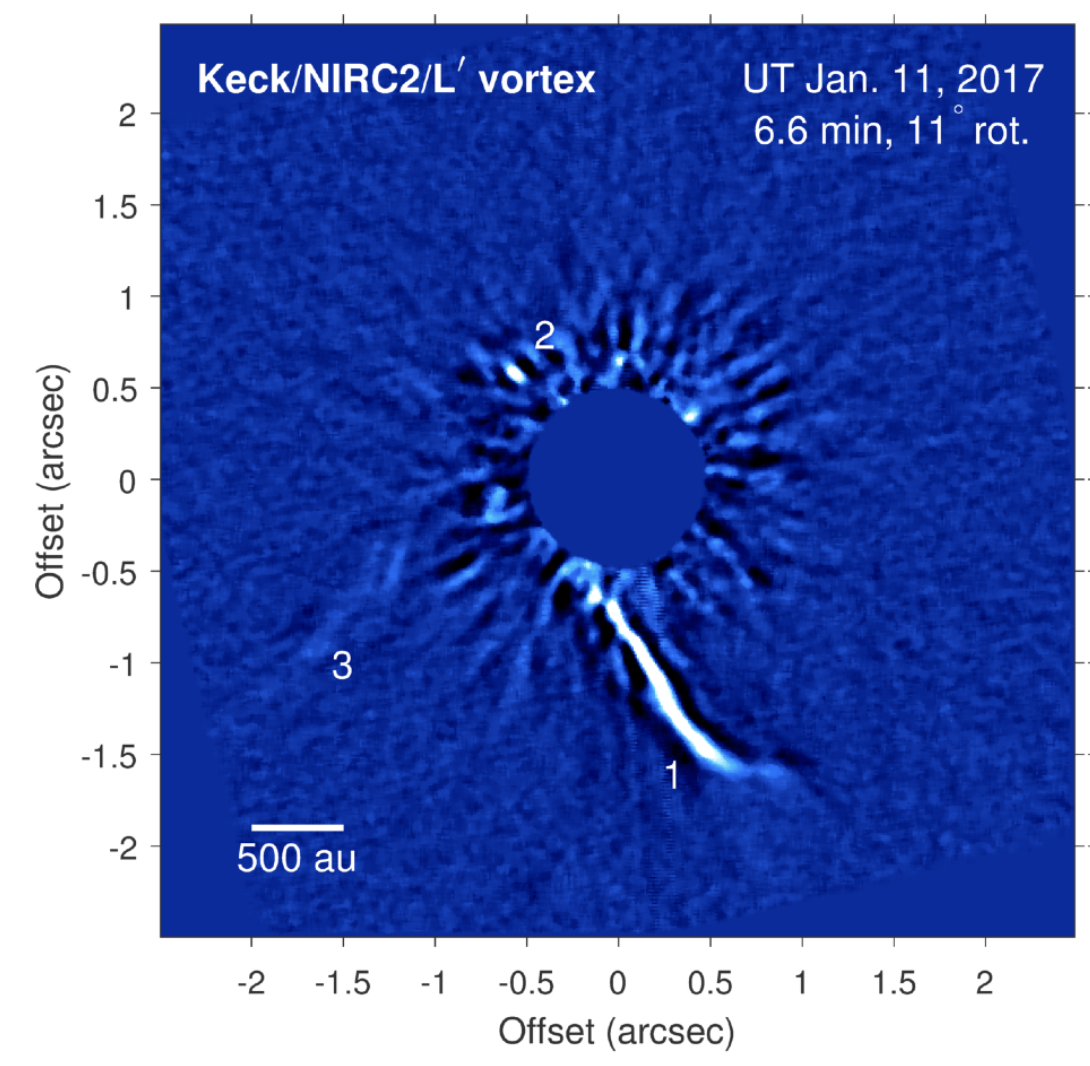}
  \caption{ADI image of Z CMa taken with Keck/NIRC2. The streamer is visible and marked as feature 1, then a point source is detected, labeled as 2, and a faint emission is also present in the southeast of the star (feature 3). }
\label{f:keck}
\end{center}
\end{figure}

{Another} point source is detected to the North-East of the central star, {also showing emission in the 1.3 mm image (see Fig.~\ref{f:super})}. To investigate the nature of this object, we reduced Keck/NIRC2 archival data of the system. The NIRC2 image was obtained on 2017-1-11 in the L' filter, employing the vortex coronagraph. The final angular differential imaging \citep[ADI,]{2006ApJ...641..556M} reduced image is shown in Fig.~\ref{f:keck}. {The quality of the data is poorer than the SPHERE images due to the low rotation angle of 11 degrees, so the image was used with the purpose of re-detecting the point source}.

To check if the point source (object 2 in Fig.~\ref{f:keck}) is comoving with the system, we measured the astrometric positions in the two epochs separated by 5 years. In the SPHERE data we measured its position with respect to the central star to be $\Delta$RA = 549 $\pm$ 3 mas and $\Delta$Dec = 555 $\pm$ 3 mas {with the same method presented in \citet{2016A&A...587A..57Z}}.  {In the NIRC2 image the measured position is $\Delta$RA = 549 $\pm$ 15 mas and $\Delta$Dec = 558 $\pm$ 15, the error bars include the standard deviation of the results of three different centering methods and the True North error. Given the proper motions of the star (in 5 years it moves $\Delta$RA = -24 mas and $\Delta$Dec = 19)}, the object is indeed very likely part of the system, which makes Z CMa a triple system and not a binary. The contrast of the star is $\Delta$H = 11.2 and $\Delta$L = 10.3 magnitudes {with respect to the star behind the coronagraph (northern component). }

\subsection{XZ Tau}
XZ Tau is a known binary system \citep[e.g.,][]{2016AstL...42...29D}, composed of an M1.5 spectral type star and a less massive (M3/M3.5) southern companion \citep{2001ApJ...556..265W}. The distance to the system is 146 pc, assuming it is part of the L1551 region \citep{2020A&A...638A..85R}. Both components of XZ Tau are spectroscopically cataloged as EXor in \cite{2004A&A...419..593C, 2022ApJ...929..129G} {where the authors reported the evolution of the outburst, compatible with EXor stars}. 

XZ Tau was identified with a bipolar collimated outflow in \cite{1990A&A...232...37M} and later spatially resolved with the HST in \cite{1997ApJ...481..447K}. A multi-epoch HST study of the XZ Tau binary using R-band, H-alpha, and [S II] emission between 1995 and 2005 shows this outflow is complex and expanding \citep{2008AJ....136.1980K}. The northern outflow appears as a succession of bubbles and a faint southern counterbubble was identified with HST data in 2004. As discussed in more detail in \cite{2008AJ....136.1980K}, while both XZ Tau A and B have stellar jets, the bubble and counterbubble appear aligned with the jet from XZ Tau A. This morphologically complex feature can be explained as previously ejected material interacting with the jet component of XZ Tau A. In 2005, the length of the northern XZ Tau bubble extended $\sim$6 arcseconds from the binary midpoint. ALMA imaging of this source confirms that the bubble features align with the XZ Tau A component \citep{2015ApJ...811L...4Z}.

The system was observed with ALMA several times, in Cycles 2, 4, and 5 \citep{2021ApJ...919...55I}. In the ALMA dust continuum emission the two components of the system, XZ Tau A (in the South-East) and XZ Tau B (in the North-West) show two circumstellar disks. Both of them are unresolved even in the highest resolution configuration (0\farcs03). \cite{2021ApJ...919...55I} estimated disk masses of $(0.77\pm3.47)\times 10^{-3}$\MSun for A and $(0.96\pm4.36)\times10^{-3}$\MSun for B. Combining the ALMA astrometry from the three Cycles and the NIR observations presented in \cite{2016AstL...42...29D}, the best orbital fitting indicates a {non-circular} orbit of the two components of XZ Tau. Finally, \cite{2021ApJ...919...55I} found that not only the two circumstellar disks are misaligned to each other, but also to the orbital plane of the binary system. They suggest that this is an indication that turbulent fragmentation played a major role in the formation of this young binary system.

\newpage


In the IRDIS FoV, the binary is resolved, and no other object is detected. An outflow is seen towards the northeastern part of the system. This faint structure was not observed before and has no counterpart in the ALMA data. The feature seems to be connected with the lower mass A component. The outflow extends for several arcseconds, towards the NE of the system. Diffuse signal is recovered in a cone shape with the axis pointing towards the North-East of the system. On the opposite side, very faint emission also suggests that we are seeing the southern outflow of the YSO. Indeed, in the same regions, gas emission is detected in the $^{12}$CO channel map \citep{2021ApJ...919...55I}.


\subsection{FU Ori}

The star FU Orionis is the archetype of the FUor eruptive objects. For this reason, it has been {extensively} studied in the past. It is located at a distance of $408\pm3\,$pc \citep[Gaia DR3;][]{2021AJ....161..147B} and has a mass of 0.3 \MSun \citep{2012AJ....143...55B}. The age of the system is estimated to be $\sim$ 2 Myr \citep{2012AJ....143...55B}. FU Ori is part of a wide binary system, its companion FU Ori S, probably a 1.2 \MSun pre-main sequence K star, is located at a separation of $\sim$ 0\farcs5 away and position angle of 161 degrees \citep{2004ApJ...601L..83W,2012AJ....143...55B}. FU Ori started its outburst in 1936 {\citep{1954ZA.....35...74W,1966VA......8..109H}}.

The surroundings of both stars in the binary system present various features: using Subaru/HiCIAO NIR imaging, \cite{2016SciA....2E0875L} detected a bright arm, or arch-shaped feature, east of the primary star. The same finding was presented in \cite{2018ApJ...864...20T} using polarimetry with HiCIAO, together with a tail at 3 arcsec ($\sim$1300au) to the west. Using Gemini/GPI, again with polarimetry, \cite{2020ApJ...888....7L} found the same structures, with the addition of a dark stripe toward the northern tip of the eastern arm, and a diffuse region of scattered light towards the SW. The tail appears to be very faint in the GPI images.

{Early ALMA  observations provided the first detection of CO gas surrounding the FU Ori binary system \citep[at 0\farcs6 $\times$ 0\farcs5 resolution;][]{2015ApJ...812..134H}. Higher-angular resolution observations suggest an interesting correspondence between the CO gas and the arm-like feature identified in previous HiCIAO and our novel SPHERE observations \citep{2020ApJ...889...59P}, as expected for micron-sized dust particles well-coupled to the gas.}

Note that a first SPHERE DPI epoch on FU Ori from 2016 has been presented in \cite{2023MNRAS.518.5620W}. In this work, we present a second epoch, acquired in 2017, shown in the gallery in Fig~\ref{f:mos}. A comprehensive analysis comparing the two epochs will be the subject of a future publication.

\subsection{V960 Mon}
V960 Mon is an FUor object \citep{2023ApJ...945...80C} located at a distance of {2189$_{-249}^{+321}$ pc} \citep{2021A&A...649A...1G}. This object has been in an outbursting phase since 2014 {\citep{2014ATel.6797....1H}}. The eruptive star has a stellar companion located at a separation of 0\farcs2 \citep[see, e.g.][]{2015ApJ...806L...4C}. This spectacular system, including the SPHERE data, is presented in a dedicated publication, \citet{2023ApJ...952L..17W}, and we refer the reader to that work for any details on the source. The authors presented the first evidence of planetary-mass clumps along the spiral of this FUor object. V960 Mon shows a complex structure of several spirals that extend for thousands of au.  For completeness, we show in Fig.~\ref{f:mos} the SPHERE image of the system.

\subsection{UZ Tau}
The system of UZ Tau is a quadruple, composed of a spectroscopic binary, UZ Tau E, with a separation of $\sim$0.03 au \citep{1996AAS...188.6005M} and surrounded by a large circumbinary disk \citep{2018ApJ...861...64T}, and an M3+M3 binary pair with a 48 au projected separation, UZ Tau W. {UZ Tau E} is located at a distance of $123\pm7\,$pc \citep{2023A&A...674A...1G}, the RUWE value is high (15.3). The {EXor} object is the spectroscopic binary UZ Tau E, which shows moderate short-term variability typical of the EXor objects \citep{1977ApJ...214..747H, 2007ApJ...665.1182L}. Its variability could be explained by the interaction between the spectroscopic binary and the disk, as suggested in \citep{2007AJ....134..241J}. 

UZ Tau E was observed with ALMA by \cite{2018ApJ...869...17L} and \cite{2020ApJ...900....7H}. \cite{2018ApJ...869...17L} observed the source with a 0\farcs13 $\times$ 0\farcs11 resolution, and they reported an inner cavity and two emission bumps. They found three rings at 11, 17, and 77 au from the central stars. From the $^{13}$CO and C$^{18}$O emission, \cite{2019ApJ...883...22C} determined that the circumstellar disk is almost coplanar with the orbit of the spectroscopic binary. \cite{2020ApJ...900....7H} observed again the system with a lower resolution and derived the mass of the disk (22 \MJup), the inclination (59 deg), the major and minor axes (668 and 396 mas, respectively). The disk shows a clear {Keplerian} pattern of a Class II disk. This might be an indication that the EXor objects are in a more evolved stage than FUors.

The reduced IRDIS image is shown in Fig.~\ref{f:mos}. An inclined disk is detected around UZ Tau E, even if the emission is faint. No substructure is seen in the NIR. No extended emission around the system is detected in the IRDIS image. A larger FoV is also shown in Fig.~\ref{f:bin} and presented in Sec.~\ref{s:multi}, where the visual binary system UZ Tau W is detected.

\subsection{NY Ori}
NY Ori, also known as Parenago 2199, is a young eruptive star located at a distance of $403\pm5\,$pc \citep{2021A&A...649A...1G} in the Orion Nebula Cloud complex, {more specifically in the Orion A subcloud,} \citep{2019A&A...628A.123Z}. In \cite{2022ApJ...929..129G} it is listed as a single system, even if it is visually close to the star V566 Ori (also called Parenago 2118), located at 5.5 arcsec from NY Ori, and a distance of {$388\pm3\,$ pc} \citep{2020yCat.1350....0G}. This close star polluted many light curves of the EXor object in the past \citep[see, e.g,][]{2017A&A...602A..99J}. NY Ori was part of the ALMA FUor/EXor survey presented in \cite{2018MNRAS.474.4347C}. They found that the continuum emission shows a single disk, spatially resolved, with a mass of 40$^{+10}_{-10}$ \MJup and radius 76$^{+5}_{-7}$ au {inferred for the radiative transfer modeling, which assumes a gas-to-dust ratio of 100}. The disk is also detected in the gas lines $^{13}$CO and C$^{18}$O, as a compact emission. There is no evidence of outflows from the $^{12}$CO emission. \cite{2017A&A...602A..99J} reported that no significant outburst (with a duration longer than 6 months) occurred during monitoring of 40 yr (since 1961). On the other hand, \cite{2016ATel.8828....1L} reported on a rapid brightness variability monitored from 2011 to 2015 in a very short cadence (e.g., a fading of 1.4 mag in 2 days). The monitoring was part of the EXORCISM program \citep[EXOR OptiCal and Infrared Systematic Monitoring,][] {2013prpl.conf2B055A, 2009ApJ...693.1056L}. 

There are no other {high-contrast} visible/NIR data in the ESO or Keck archives besides the SPHERE data presented in this work. In the IRDIS image, the signal around the EXor is very low, but a faint disk is seen around the star. The reduced image is shown in Fig.~\ref{f:mos}. The close star V566 Ori is {easily} visible in the NW of NY~Ori, as shown in Fig.~\ref{f:bin}. The astrometrical positions are: $\Delta$RA = 3\farcs280 $\pm$ 0\farcs006, $\Delta$Dec = 4\farcs711 $\pm$ 0\farcs004  with respect to the central star.

\subsection{EX Lup}
EX Lup is the prototype of the EXors objects \citep[see, e.g.,][]{2009A&A...507..881S,2010ApJ...719L..50A}. The IRDIS/SPHERE data are extensively presented in \cite{2020A&A...641A..33R}. We include the final image in Fig.~\ref{f:mos} for completeness of the sample. The detection of the close environment around EX Lup is very similar to the one around NY Ori. 

\section{Discussion}
\label{s:disc}

\begin{figure*}[h!]
\begin{center}
\includegraphics[width=0.95\textwidth]{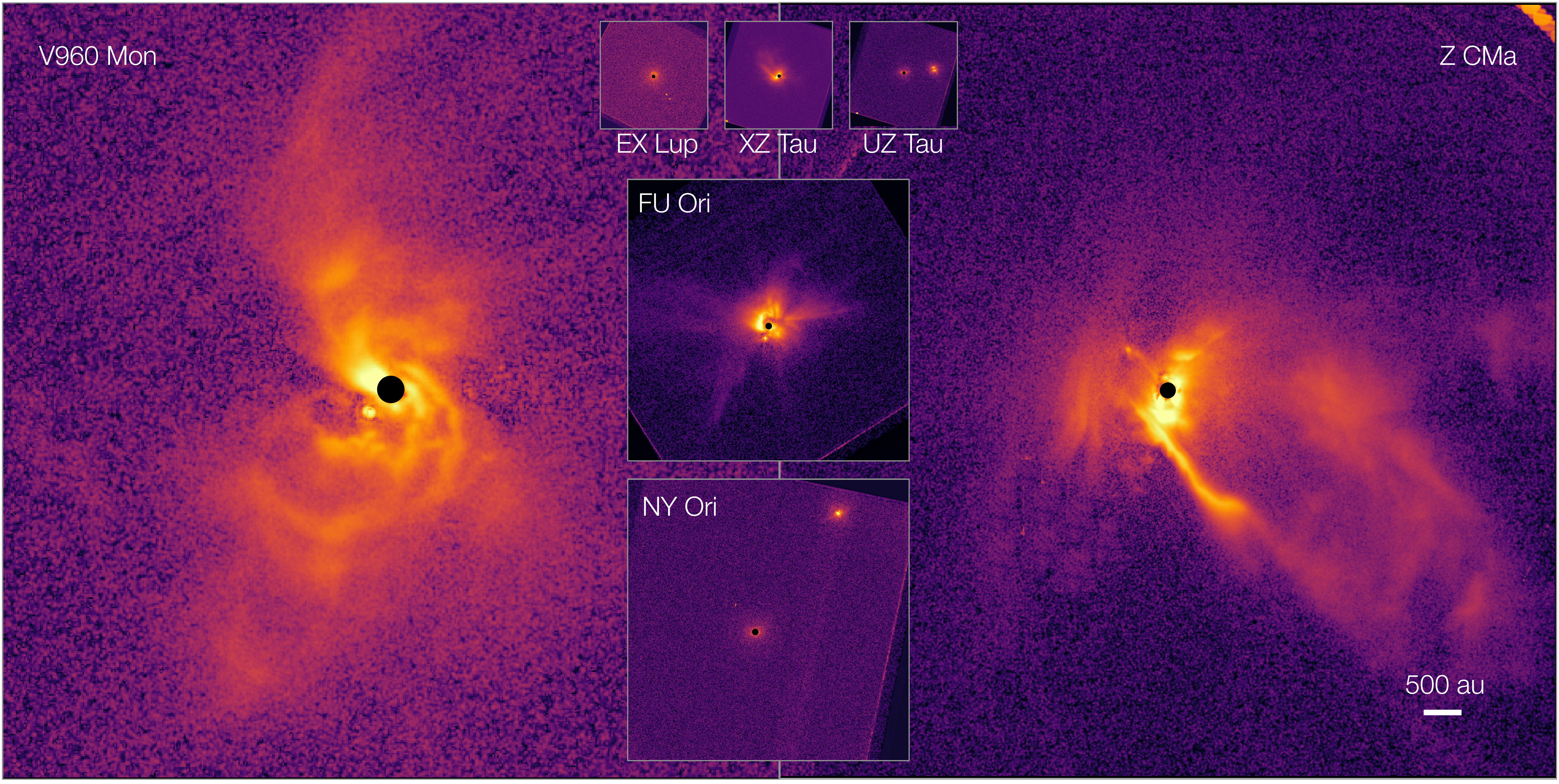}
  \caption{The SPHERE polarized light images of the eruptive stars {at the same projected spatial scale}. {The bar marking 500$\,$au is shown in the bottom right corner and applies for all seven sub-set images.}}
\label{f:mos_scale}
\end{center}
\end{figure*}

\subsection{Taxonomy of eruptive stars}
Despite the small number of eruptive stars that are observable in polarized light, it is still possible to discern between 3 types of structures that are found around FUor/EXors: objects with outflows, asymmetric arms, and faint disks. The structures seen in these systems span a huge range of spatial scales as it can be visualized in the up-to-scale Fig.~\ref{f:mos_scale}. 

In the first category are Z~CMa and XZ~Tau. In both cases, only one side of the putative outflow is bright enough to be well detected in the SPHERE images. This is probably due to an illumination effect and tells us that the part of the outflow visible in polarized light is the near side of the hourglass. The distances to these two objects differ by an order of magnitude. Consequently, the outflow seen in the southern part of Z~CMa extends {for $\sim$ 6000 au}, making it more than 10 times bigger than the outflow on the Nothern side of XZ~Tau. In this context, it is important to note that Z~CMa is an FUor object, while XZ~Tau is an EXor.  

FU~Ori and V960~Mon, both FUor objects, show complex asymmetric features. For these two objects, the onset of their current outburst was recorded, in 1936 and 2014, respectively. FU~Ori shows a bright arm, arch-shaped features, and a faint tail towards the East. The structure of the environment around FU~Ori makes it unique, with no similar object among the other YSOs. V960~Mon is surrounded by huge spiral arms, extending for thousands of au, both in the South and in the North. V960~Mon and Z~CMa show the most extended structures of the sample. 

For UZ~Tau, NY~Ori, and EX~Lup, all labeled as EXors, the signal is similar to that expected from self-shadowed disks \citep[see, e.g.,][]{2022A&A...658A.137G}. UZ~Tau is surrounded by an inclined and faint disk, NY~Ori is only marginally resolved and shows an asymmetry on the southwest. Very similarly, EX~Lup also has an inclined disk, very faint and compact. In summary, among the EXors of the survey, only XZ~Tau shows a prominent feature, while the others show mostly axisymmetric faint disks.

{There is no obvious reason why an inclined disk should be faint in scattered light. Quite the opposite, on average they appear brighter because of smaller angles of scattering, providing an increased number of photons, and are probed \citep{2022A&A...658A.137G}. Therefore, the faintness of the disks in question must be intrinsic to the system.}

It is important to note that the three FUors are presently all in the range $L_{bol}$ of hundreds of \LSun; the EXors are a few tenths to a few \LSun \citep[see Table 1 in][]{2014prpl.conf..387A}. This luminosity dichotomy can explain the different illumination of the extended features that we see around the eruptive stars and their luminosities.

\subsection{Polarization analysis}
The polarization of light observed from an object or area can be described by its Degree of Linear Polarization ($DoLP$) and the Angle of Linear Polarization ($AoLP$). Those two quantities give the polarization fraction and direction and are related to the Stokes components:
\begin{eqnarray}
    {DoLP} &=& \frac{PI}{I}\,,\\
    {AoLP} &=& \frac{1}{2}\arctan\left(\frac{U}{Q}\right)\,,
\end{eqnarray}
where $I$ is the total intensity.

For bright objects, locally dominating the total intensity, one can calculate both $DoLP$ and $AoLP$. Therefore, we have quantified the polarization contained within the signal of all distinguishable stars in our observational dataset. The corresponding values are listed in Table~\ref{t:stellar_pol}.
It is important to note that the commonly used term `stellar polarization' can be somewhat misleading, as the observed polarization of a star does not represent an inherent quality of its emitted light, but rather arises from unresolved scattering within the point-spread function of the telescope. 
This scattering can originate from dust grains close enough to the star to not be discernible, or it may be influenced by scattering events along the line of sight, potentially involving interstellar dust particles \citep[see][for a discussion]{2021A&A...647A..21V}.
For the systems that do not show significant scattering in the region of the residuals of the adaptive optics (UZ~Tau, NY~Ori, EX~Lup), we measured the central star's polarization from a mask over this area, as directly provided by the IRDAP pipeline. 
For the other systems and the off-centered companions, we measured the polarization directly from an aperture of 8 pixels centered on the star. For the measurement we reject the pixels that are closer than $0\farcs05$ to the star to avoid including saturated pixels.
The values' systematic variability when varying the shape and size of this mask is expected to dominate the error of the measurements.
\begin{table}
\begin{minipage}{0.5\textwidth}
\caption{Stellar Polarizations expressed by the Degree of Linear Polarization ($DoLP$) and Angle of Linear Polarization ($AoLP$).} 
\label{t:stellar_pol}
\renewcommand{\footnoterule}{}  
\centering
\begin{tabular}{lrr}
\hline
\hline\\[-0.3cm]
  Object    & $DoLP$    &  $AoLP$   \\
           \hline\\[-0.28cm]
            Z CMa &  0.3\%  & 70$^\circ$ \\
            Z CMa `D' &  1.1\%  & 85$^\circ$ \\
            XZ Tau A& 2.6\%  & 31$^\circ$ \\ 
            XZ Tau B& 3.1\%  & 154$^\circ$ \\ 
            FU Ori N& 0.3\%& 86$^\circ$\\
            FU Ori S& 0.4\% & 128$^\circ$ \\
            V960 Mon A& 1.7\% & 107$^\circ$\\
            V960 Mon B& 0.2\% & 142$^\circ$ \\
            UZ Tau E & 0.7\%  & 2$^\circ$ \\ 
            UZ Tau Wn & 0.5\%  & 37$^\circ$ \\
            UZ Tau Ws & 0.3\%  & 48$^\circ$ \\
            NY Ori& 0.4\% & 33$^\circ$  \\
            Parenago~2118$\footnote{Parenago~2118 was measured from the FoV around NY~Ori.}$& 0.4\% & 118$^\circ$   \\
            EX Lup & 0.5\%& 168$^\circ$  \\
\hline
\end{tabular}
\end{minipage}
\end{table}

\begin{figure*}[h!]
\begin{center}
\includegraphics[width=0.7\textwidth]{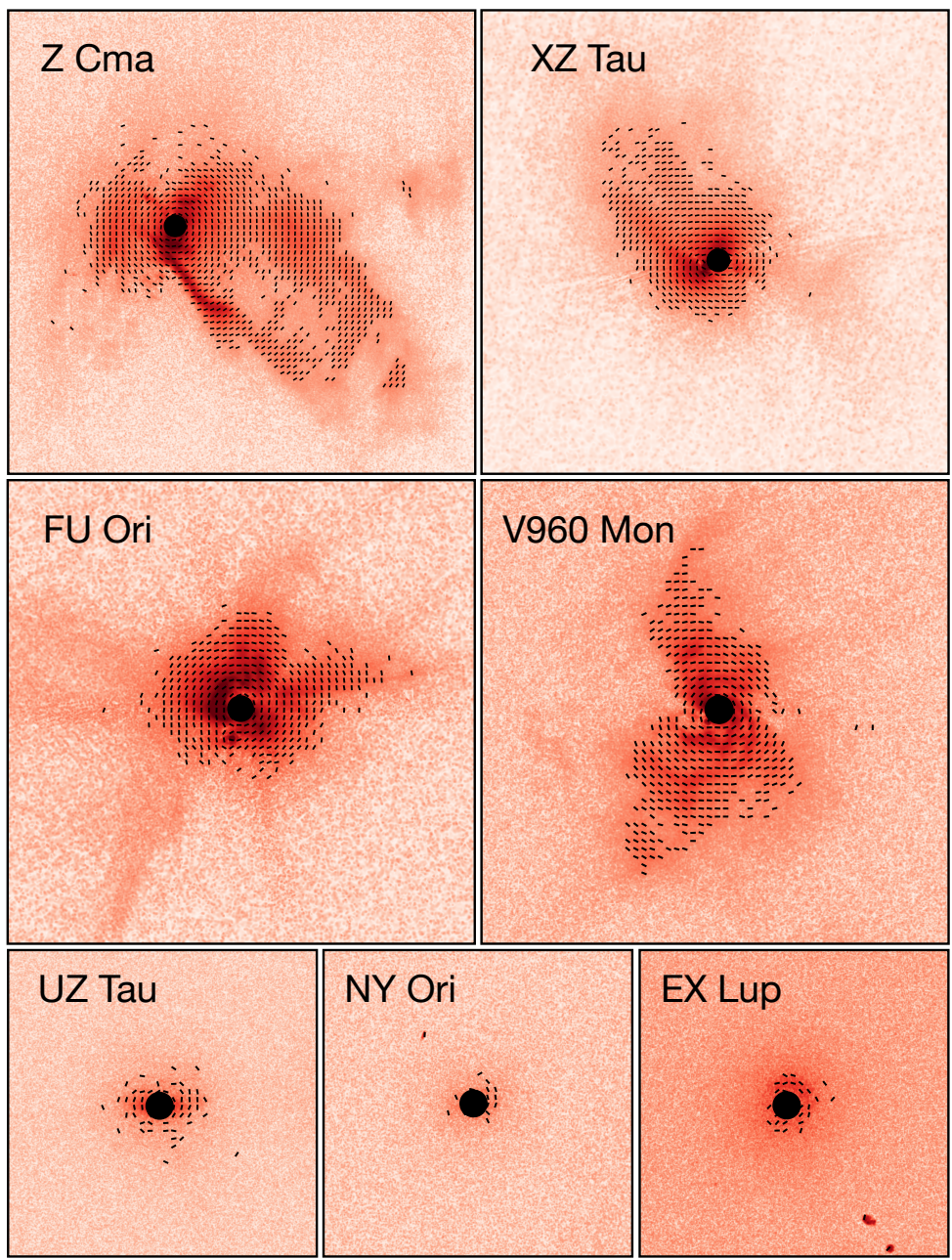}
  \caption{Polarization vectors measured in the IRDIS images. We only plot the vectors where the signal-to-noise is high enough to allow for a reliable assessment.}
\label{f:mos_polvec}
\end{center}
\end{figure*}

For circumstellar scattering regions, it is often impractical to determine the $DoLP$ when the total intensity is predominantly influenced by the stars themselves, while the polarized intensity is a result of light scattering off circumstellar material.
Consequently, our analysis is confined to the measurement of the $AoLP$. Figure~\ref{f:mos_polvec} presents an adapted version of the polarized light gallery from Figure~\ref{f:mos}, with an emphasis on the superimposed polarization vectors that represent the local $AoLP$. Under conditions involving a single light source and singular scattering events, these vectors are anticipated to align orthogonally to the direction of incoming light. As shown in Figure~\ref{f:mos_polvec}, the observed vector patterns suggest illumination predominantly by the central star. However, for NY~Ori and EX~Lup, definitive conclusions are challenged by the low signal-to-noise ratio in the data.

\subsection{Reinterpretation of Z~CMa's streamer}

%

Our observations reveal a significant connection between the streamer in Z~CMa and an extensively illuminated large-scale outflow-like structure that was previously undetected. The consistent opening angles between the streamer and the illuminated outflow suggest the need for a reevaluation of the recently proposed tidal tail interpretation \citep{2022NatAs...6..331D}.

It is worth noting that the mm-emission at the end of the tail \citep[a proposed stellar flyby, marked as ``C'' in their Fig.~1;][]{2022NatAs...6..331D} does not correspond to any visible counterpart in either the total intensity or polarized intensity images of the SPHERE observation as in Fig.~\ref{f:super}.
Furthermore, recent findings related to possible gravitational instability clumps around the erupting object V960 Mon provide valuable evidence for clumps of mm-emission found at intermediate distances around eruptive stars \citep{2023ApJ...952L..17W}. uture ALMA observations of the source may shed light on the nature of the emission, which might be an ejected remnant, possibly linked to the clumps that triggered Z~CMa's eruptive behavior.

Nevertheless, the exceptional collimation exhibited by the streamer-like feature on one side of the extended outflow, as seen in polarized light, coupled with its distinctive hook-like morphology at its tip, poses a challenge to a typical outflow model. One possibility could be that the ALMA `C' source identified as a flyby object may not have directly perturbed the circumstellar disk but, instead, may have perturbed the large-scale outflow, resulting in the collimation of material to form the observed elongated filament. Investigating this idea through hydrodynamic flyby simulations, while considering the influence of the primary star's envelope and outflow, could offer valuable insights into the collimation of material and its connection to dynamical effects, where a massive object in motion leaves behind a collimated wake.

\subsection{Multiplicity}
\label{s:multi}
\begin{figure*}
\begin{center}
\includegraphics[height=0.45\textwidth]{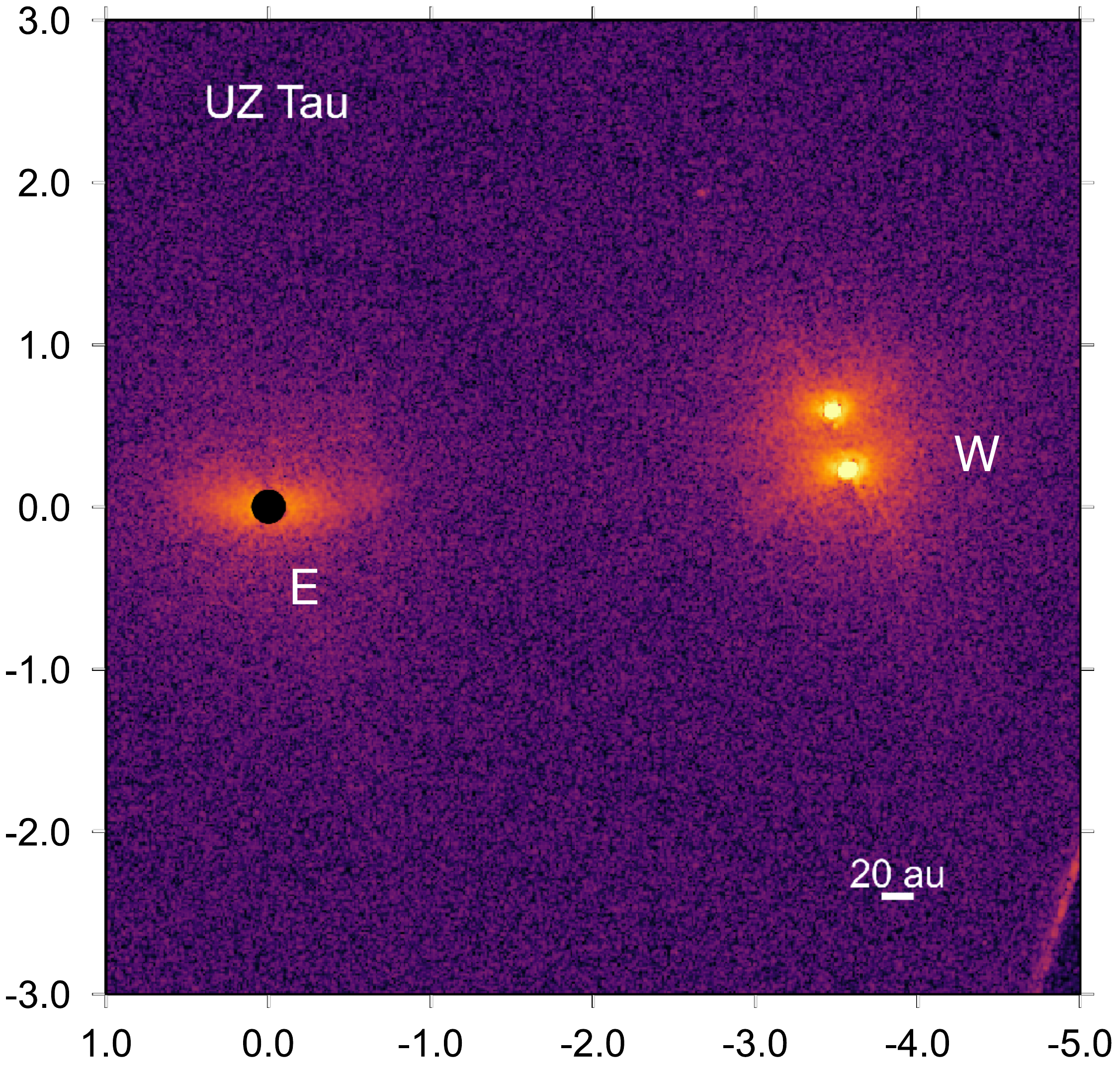} \hfill
\includegraphics[height=0.45\textwidth]{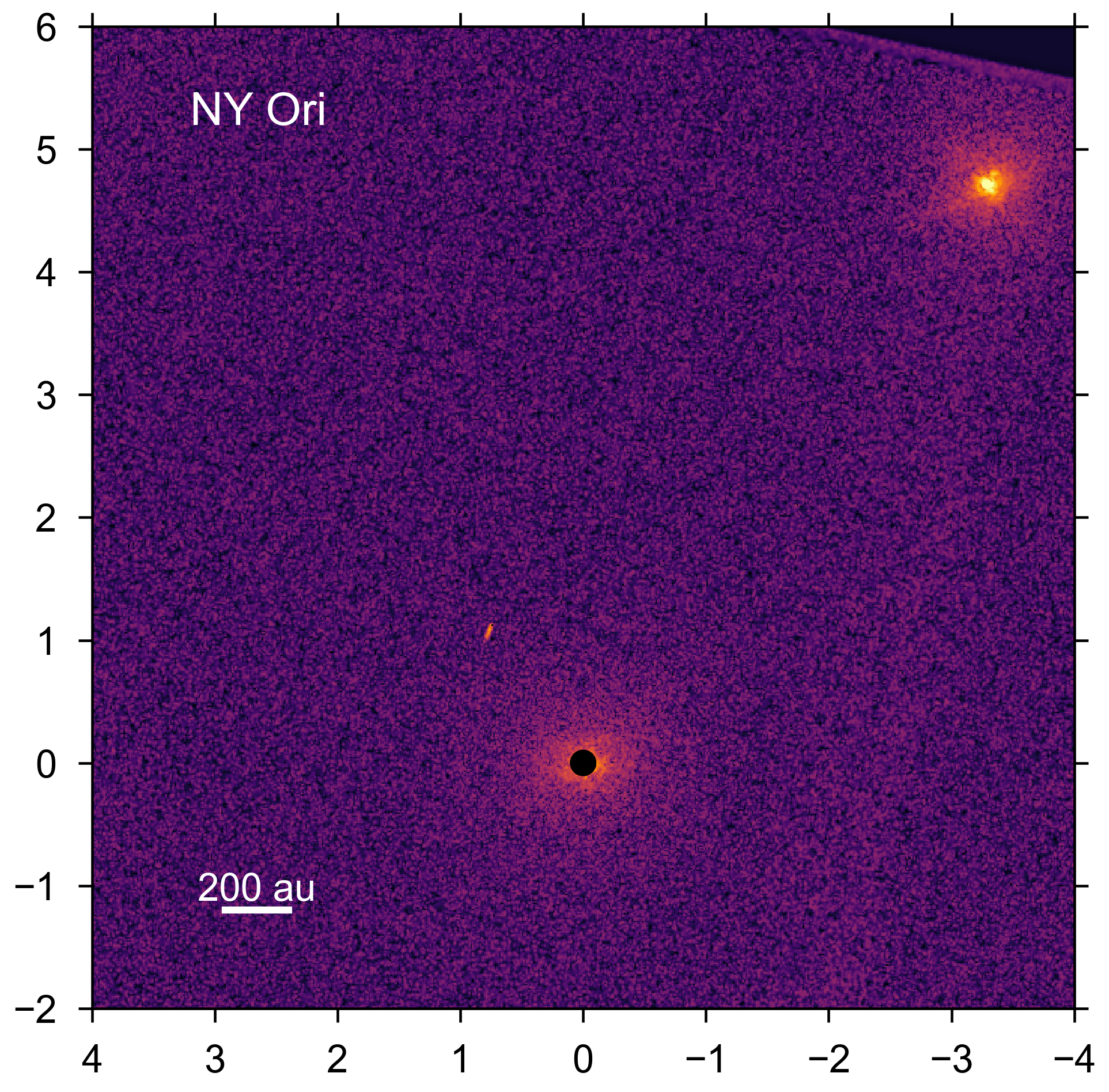}  \\
\caption{Visual binaries around the systems of UZ~Tau and NY~Ori in the polarized intensity images.}
\label{f:bin}
\end{center}
\end{figure*}

Among the stars in our sample, 5 out of 7 systems are multiple. Most probably NY Ori and EX Lup are the only single stars, even if Parenago 2118, the star located to the northwest of NY Ori share a similar parallax (see Fig.~\ref{f:bin}). In the case of EX~Lup, {the data presented here exclude the presence of a visual binary, and} \cite{1998MNRAS.301..161B} excluded a spectroscopic companion. On the other hand, \cite{2020A&A...641A..33R} {proposed that the one-sided polarized light signal around EX~Lup could be due to shadowing by a moderately misaligned inner disk. In a general study, \citet{2019MNRAS.484.4951N} showed that the physical origin of this misalignment can be due to a massive planet on an inclined orbit}. In general, observations of young circumstellar disks demonstrated that gas and dust surrounding YSOs can be heavily affected by multiplicity \citep[see, e.g.][and references therein]{2014ApJ...784...62A, 2019ApJ...872..158A,2020MNRAS.496.5089Z,2021MNRAS.501.2305Z, 2023EPJP..138..411Z, 2023ASPC..534..275O}. 

The fact that multiplicity can be related to eruptive stars was suggested in the past \citep[see e.g.][]{1992ApJ...401L..31B}. \cite{2004IAUC.8441....3R} suggested that the decay of a binary system due to the interaction of the companion with the circumbinary disk can cause disk instability.
With the exception of the two objects mentioned above, the systems presented in this survey are all visual multiples: Z~CMa has two stellar companions at $\sim$100$\,$ and $\sim$800$\,$au separation (detected in our data for the first time), XZ~Tau at $\sim$30$\,$au, FU~Ori at 200$\,$au, V960~Mon at 400$\,$au, and UZ~Tau is a quadruple system. Note that, even if it seems that most eruptive objects have companions, the majority of these multiple systems are widely separated. As pointed out by \cite{2014prpl.conf..387A}, if binarity was the main trigger of the outburst, it would happen preferably in close binary systems (i.e. less than 10$\,$au). Apart from UZ~Tau, which is a spectroscopic binary, {this is not known to be the case} for all the other systems. To confirm or discard this mechanism as the origin of eruptive stars high-resolution/spectroscopic observations are needed to probe the stars' vicinity. In particular, a high-contrast imaging campaign to look for giant companions around the stars would help in understanding if sub-stellar close companions are responsible for the outbursts, as suggested in \cite{2004MNRAS.353..841L}.  

\section{Conclusions}
\label{s:con}

In this paper, we report on a survey of seven eruptive stars observed in polarized light with the instrument SPHERE/IRDIS. The survey aimed at observing all the episodically accreting objects observable with the instrument SPHERE, whose adaptive optics system is limited by the magnitude of the central star in the visible. Given the young age of these kinds of objects, most of them are embedded and therefore faint in the visible. The survey observed Z~CMa, XZ~Tau, FU~Ori, V960~Mon, UZ~Tau, NY~Ori, and EX~Lup. Three objects are FUor-like sources (Z~CMa, FU~Ori, and V960~Mon), and the rest are EXors. All the objects show polarized signals in the IRDIS images. 

Our observations of eruptive stars in polarized light reveal 3 main structures: outflows, asymmetric arms, and faint disks. Notably, Z~CMa and XZ~Tau represent the outflow category, with Z~CMa's extensive outflow surpassing XZ~Tau's. It is worth noticing that Z~CMa is an FUor-like object, while XZ~Tau is an EXor. FU~Ori and V960~Mon, both FUor objects, exhibit distinct features, while UZ~Tau, NY~Ori, and EX~Lup, identified as EXors, display faint disk-like signal, with XZ~Tau standing out for its prominent feature. Being younger and far more luminous, the FUors drive and illuminate outflows, whereas the EXors no longer drive outflows (at least, not extensive FUor-like outflows) and would be incapable of illuminating the outflow walls/cavities to the same extent as the FUors.

The unprecedented quality of the polarized light image for Z~CMa revealed faint large-scale structures that seem to be connected to a previously detected, bright filament. This challenges the interpretation of this filament as a fly-by feature and suggests its connection to a past outflow event.

Among the stars observed, the majority are multiple systems, except for NY~Ori and EX~Lup. Evidence indicates that the interaction of companions with the circumstellar disk could lead to disk instability, possibly triggering the outburst in eruptive stars. In fact, the systems surveyed here have various separations, the presence of {\it widely} separated companions might not be the primary trigger for the outbursts. However, further high-resolution observations, especially in terms of high-contrast imaging campaigns in pupil-tracking mode, are needed to investigate the role of giant or substellar {\it close} companions in the eruption mechanisms of these stars. For example, the possibility of a substellar companion shaping the disk around EX Lup was proposed in the past. SPHERE, once again, is a good instrument to perform this kind of observation, along with GPI, SCExAO, or JWST.

\begin{acknowledgements}
We thank the two referees for their comments that substantially improved the manuscript. We are thankful to R. Dong for the fruitful discussions and for providing the ALMA data on Z CMa. The authors acknowledge support from ANID -- Millennium Science Initiative Program -- Center Code NCN2021\_080. The National Radio Astronomy Observatory is a facility of the National Science Foundation operated under cooperative agreement by Associated Universities, Inc. P.W. acknowledges support from FONDECYT grant 3220399. S.P. acknowledges support from FONDECYT Regular grant 1231663. 
\end{acknowledgements}

\bibliographystyle{aa}
\bibliography{sphere_exors}
\newpage

\begin{appendix}
\section{Gallery in flux units}
We present in this Appendix the same gallery of Fig.\ref{f:mos} with a fixed color bar for all the objects. Each color indicates the surface brightness in mJy/arcsec$^2$.
\begin{figure*}[h!]
\begin{center}
\includegraphics[width=0.8\textwidth]{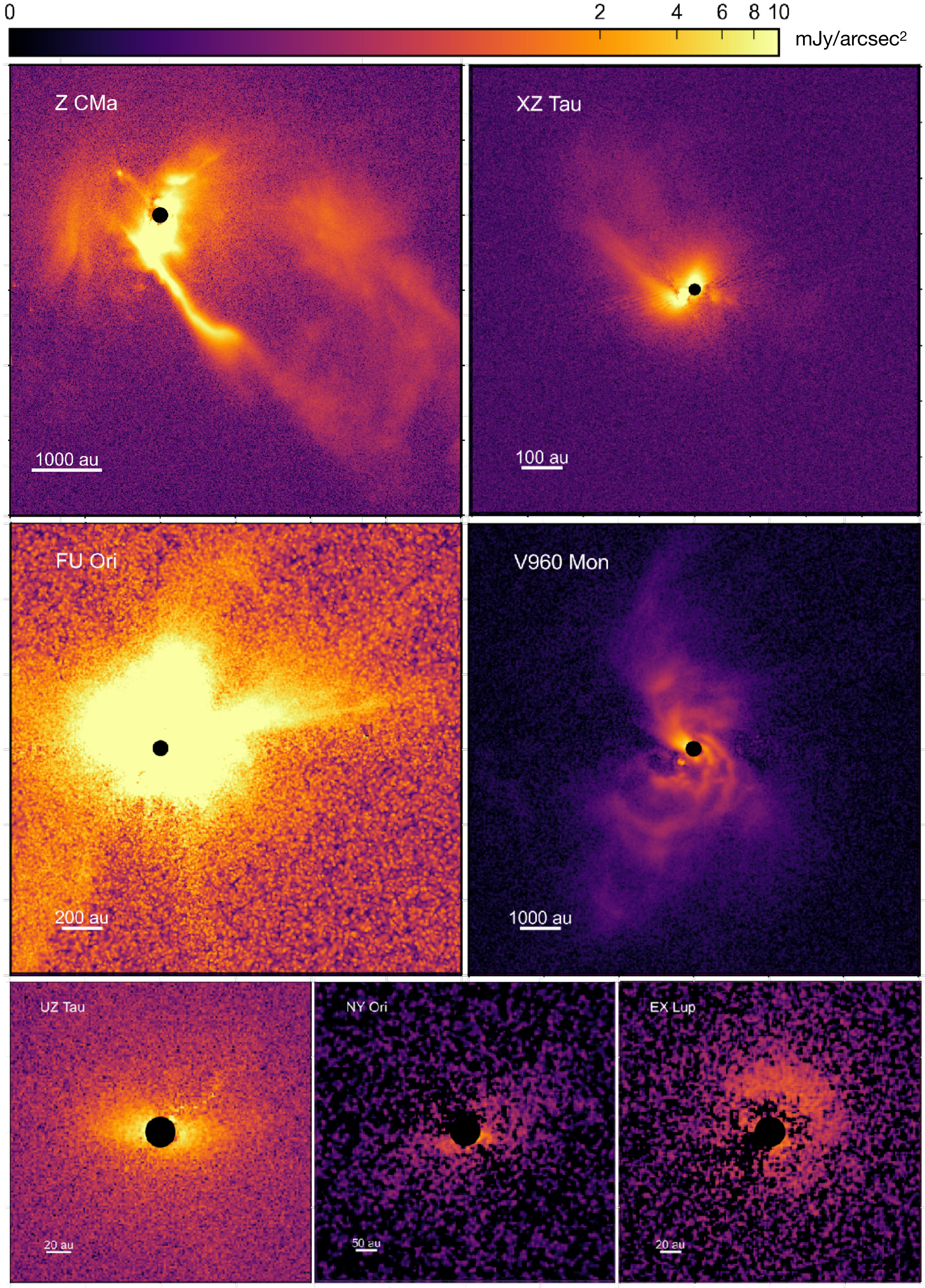}
  \caption{The SPHERE polarized light images of the eruptive stars. The color bar applies to all the images and helps the reader in retrieving the surface brightness of the features around the stars.}
\label{f:appendix}
\end{center}
\end{figure*}

\end{appendix}

\end{document}